\documentstyle[11pt,aaspp4]{article}

\newcommand{\kms}{\, {\rm km\, s}^{-1}}
\newcommand{\cm}{\, {\rm cm}}

\newcommand{\mpc}{\, {\rm Mpc}}

\newcommand{\lya}{Ly$\alpha$ }
\newcommand{\etal}{et al.\ }

\newcommand{\dv}{\Delta v}
\newcommand{\dvv}{\Delta {\mathbf v}}
\newcommand{\dt}{\Delta \theta}
\newcommand{\dz}{\Delta z}
\newcommand{\pa}{\parallel}
\newcommand{\pe}{\perp}
\newcommand{\txi}{{\tilde \xi}}

\begin{document}

\title{Measuring the Cosmological Geometry from the Lyman Alpha Forest
       along Parallel Lines of Sight}

\author{Patrick McDonald and Jordi Miralda-Escud\'e$^{1}$}
\affil{ Univ. of Pennsylvania, Dept. of Physics and Astronomy}
\authoremail{ pmcdonal@student.physics.upenn.edu, 
              jordi@llull.physics.upenn.edu}
\authoraddr{David Rittenhouse Laboratory, 209 S. 33rd St., Philadelphia, PA
            19104}	      
\affil{$^{1}$ Alfred P. Sloan Fellow}

\begin{abstract}

We discuss the feasibility of measuring the cosmological geometry using the
redshift space correlation function of the \lya forest in multiple lines of
sight, as a function of angular and velocity separation. The geometric
parameter that is measured is $f(z)\equiv c^{-1} H(z) D(z)$, where $H(z)$ is
the Hubble constant and $D(z)$ the angular diameter distance at redshift $z$.   
The correlation function is computed in linear theory. We describe a method
to measure it from observations with the Gaussianization procedure of Croft
\etal (1998) to map the observed \lya forest transmitted flux to an
approximation of the linear density field. The effect of peculiar velocities on
the {\it shape} of the recovered power spectrum is pointed out. We estimate the
error in recovering the $f(z)$ factor from observations due to the variance in
the \lya absorbers. We show that $\sim 20$ pairs of quasars 
(separations $<3'$) are needed to distinguish a flat $\Omega_0=1$ 
universe from a
universe with $\Omega_0=0.2$, $\Omega_\Lambda=0.8$. A second parameter that is
obtained from the correlation function of the \lya forest is $\beta \simeq
\Omega(z)^{0.6}/b$ (affecting the magnitude of the peculiar velocities), where
$b$ is a linear theory bias of the \lya forest. The statistical error of $f(z)$
can be reduced if $b$ can be determined independently from numerical
simulations, reducing the number of quasar pairs needed for constraining
cosmology to approximately six. On small scales, where the correlation function
is higher, $f(z)$ should be measurable with fewer quasars, but non-linear
effects must then be taken into account. The anisotropy of the non-linear
redshift space correlation function as a function of scale should also provide
a precise quantitative test of the gravitational instability theory of the \lya
forest.

\end{abstract}


\section{Introduction}

  One of the methods to measure the parameters of the global
cosmological metric of the universe is to observe the angular size and
the redshift extent of a set of objects, which can be assumed to be
spherically symmetric and to follow the Hubble expansion on average
(Alcock \& Paczy\'nski 1979). More generally, this geometric factor can
be measured from a correlation function of any set of objects depending
on angular separation and redshift difference, by requiring that the
correlation is isotropic. It has long been known that this measurement
at high redshift is sensitive primarily to the cosmological constant 
(Alcock \& Paczy\'nski 1979).
However, generally the correlation of objects is due to gravitational
collapse, and the peculiar velocities make the correlation function in
redshift space anisotropic (Kaiser 1987). The effect of peculiar
velocities must be taken into account before the method can be applied.

Recently, a method has been developed to recover the power spectrum of
mass fluctuations from quasar absorption spectra, by measuring the 
one-dimensional power spectrum and converting it to the desired
three-dimensional power spectrum (Croft \etal 1998, hereafter CWKH).
This method suffers from peculiar velocity distortions similar to those
which distort the isotropy of the correlation in redshift space.
Here, we consider the accuracy in the measurement of the redshift space
correlation function from the \lya forest in nearby pairs of quasars.
We show that it is possible to disentangle the effects of geometry and
peculiar velocities, and recover the power spectrum of mass
fluctuations from the correlations in the \lya forest. We also
demonstrate that the peculiar velocities are important for correctly
deriving the shape of the power spectrum.

In \S 2 we summarize equations describing the cosmological geometry,
discuss the linear theory correlation function in redshift space,
and comment on the effect of peculiar velocities on the shape of
the power spectrum.
In \S 3 we use a random line model to estimate the error in the 
measurement of $f(z)$ from a given number of observed quasar spectra.
The discussion is given in \S 4.
Figures \ref{fitplot}-\ref{fitplot4} 
contain the main results of this paper.

\section{Linear Theory of the Lyman Alpha Forest Correlation Function}

\subsection{Cosmological Geometry}

  The following is a short summary of the cosmological equations we
shall use (these have been discussed earlier in several papers, e.g.,
Matsubara \& Suto 1996, Ballinger \etal 1997). Observations directly
measure the redshift and angular position of each object. We write the
angular and redshift separation between two objects as ($\dz$, $\dt$).  
The redshift is caused both by Hubble flow
velocities ($v_h$) and peculiar velocities along the line of sight
($v_p$). The total velocity separation along the line of
sight is $\dv_\pa = c \dz /(1+z) = \dv_h + \dv_p$. It is also
convenient to define a perpendicular velocity separation,
$\dv_\pe= c f(z) \dt$, where $f(z)$ is a dimensionless function of
redshift that includes all the dependence on the global cosmological
metric. With the assumption of isotropy, the real space two-point
correlation function of density fluctuations $\xi_r (\dv_h, \dv_\pe)$
must be a function of $\sqrt{\dv_h^2+\dv_\pe^2}$ only.
If $\xi_r$ could be measured, it would be a relatively straightforward
matter to measure $f(z)$ by simply demanding isotropy. In reality,
distances cannot be measured accurately and only the redshift space
correlation function $\xi$ can be determined, which is affected by
peculiar velocities. The peculiar velocities introduce an anisotropy in
$\xi$ of the same order as the difference in $f(z)$ between various
cosmological models.

  The quantity $f(z)$ is predicted for any cosmological model. If the
present density of matter (in units of the critical density) is
$\Omega_0$, and considering also a negative pressure component with
density $\Omega_\Lambda$ and equation of state $p=w \rho$
(the case $w=-1$ is the cosmological constant), we have,
for an open universe,
\begin{equation}
f(z)=\frac{E(z)\sinh\left[\sqrt{\Omega_R}\int_0^z (dz/E(z))\right]}
       {(1+z)\sqrt{\Omega_R}} ~,
\end{equation}
and for a flat universe,
\begin{equation}
f(z)=\frac{E(z)\int_0^z (dz/E(z))}{(1+z)} ~.
\end{equation}
Here,
$\Omega_R=1-\Omega_0-\Omega_{\Lambda}$ ($\Omega_R=0$ for a flat 
universe),
and
\begin{equation}
E(z)=\sqrt{\Omega_0(1+z)^3+\Omega_R(1+z)^2+
\Omega_{\Lambda}(1+z)^{3(1+w)}} ~.
\end{equation}
\begin{figure}
\centerline {
\epsfxsize=4.7truein
\epsfbox[70 32 545 740]{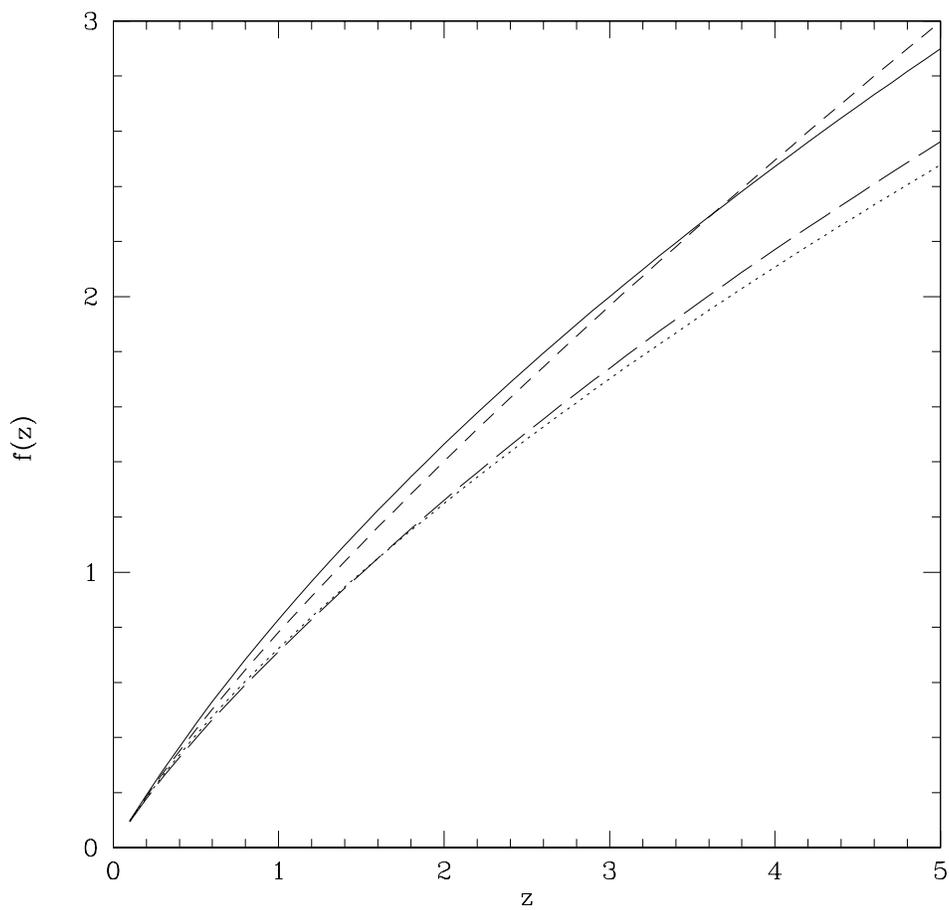}
}
\caption{$f(z)\equiv \dv_\pe/(c \dt)$ computed for $\Omega_0=1.0$,
$\Omega_\Lambda=0.0$ (solid line), $\Omega_0=0.4$,
$\Omega_\Lambda=0.6$, $w=-1$ (long dashed line), $\Omega_0=0.3$,
$\Omega_\Lambda=0.0$ (short dashed line), and $\Omega_0=0.4$,
$\Omega_\Lambda=0.6$, $w=-2/3$ (dotted line). }
\label{ffig}
\end{figure}
Figure \ref{ffig} shows predictions for 
$f(z)$.  

\subsection{The Correlation Function in Redshift Space}

  In linear theory, the redshift space correlation function of the
density field is given by (Kaiser 1987; Lilje \& Efstathiou 1989;
Hamilton 1992; Fisher 1995)
\begin{equation}
\xi(\dv_\pa,\dv_\pe)=\left(1+\frac{2}{3}\beta+\frac{1}{5}\beta^2
         \right)
         \xi_0(s)-\left(\frac{4}{3}\beta+\frac{4}{7}\beta^2\right)
	 \xi_2(s)P_2(\mu)+
	 \left(\frac{8}{35}\beta^2\right)\xi_4(s)P_4(\mu) ~,
\label{corfunc}
\end{equation}	   
where $s=\sqrt{\dv_\pa^2+\dv_\pe^2}$, $\mu=\dv_\pa/s$,
$P_l(\mu)$ are
the usual Legendre polynomials, and
\begin{equation}
\xi_l(s)=\frac{b^2}{2\pi^2}\int_0^\infty dk k^2 P(k)  \,
j_l\left[k s(1+z)/H(z)\right] ~.
\label{xil}
\end{equation}
The functions $j_l(x)$ are the usual spherical Bessel functions, and
$P(k)$ is the power spectrum of the mass fluctuations. The parameter
$\beta$ is related to the linear theory bias $b$ by $\beta = b^{-1}\,
H(z)^{-1}\, (dD/dt)/D$, where $D$ is the linear growth factor, $H(z)$
is the Hubble constant and $t$ is the age of the universe 
[see Peebles 1993; $H(z)^{-1}\, (dD/dt)/D \simeq \Omega(z)^{0.6}$ is
a good approximation in most models].

Equation (\ref{corfunc}) is valid only in linear theory, and the
correlation function $\xi$ could only be determined directly from
observations if the linear density field, $\delta$, were known. Only
the fraction of the flux that is transmitted, $F$, can be determined
along a line of sight from the \lya forest spectrum in a quasar (we
assume here that the quasar continuum has been fitted to a model,
allowing the transmitted flux fraction $F$ to be measured at every
pixel). We need a way to recover the linear density field $\delta$
from the observed $F$. In general, this cannot be done exactly since
$F$ is only known along a line of sight, and besides the chaotic nature
of non-linear evolution limits the accuracy to which $\delta$ can be
recovered. Here, we adopt the Gaussianization procedure of CWKH (see
also Weinberg 1992). Gaussianization 
assumes that the initial mass density fluctuations were Gaussian
random and that evolution approximately preserves the rank order of
densities, even as it distorts the distribution from a Gaussian.
In the case of the \lya forest, values of the observed flux decrement
are mapped monotonically onto a new variable $\delta$ required to have
the probability distribution function
\begin{equation}
P(\delta)=\frac{1}{\sqrt{2 \pi \xi(0)}} \,
          \exp\left[-\frac{1}{2}\frac{\delta^2}{\xi(0)}\right] ~.
	  \label{PDF}
\end{equation}
Of course, the fluctuation amplitude $\sqrt{\xi(0) }$ is not recovered
by the assumption that the rank order is preserved. Computing the
correlation function of the Gaussianized spectrum yields only the
ratio $\txi(\dv, \dt) \equiv \xi(\dv, \dt)/\xi(0) $.

  Once the new variable $\delta$ is obtained, the correlation function
may be computed from observations through the usual estimator given by
its definition: $\xi(\dvv) = \, < \! \delta({\mathbf v})\,
\delta({\mathbf v} + \dvv ) \! > \, $, where
$\dvv$ symbolizes the vector separation $(\Delta v_\pa ,
\Delta v_\pe )$. The average is taken over all pairs of pixels separated 
by $\dvv$ in the spectra available. However, this is not
necessarily the best estimator of $\xi$, and in general it should be
better to examine the full 2-point joint probability distribution
function. One of the main reasons for this is that the assumption that
the density rank order is preserved should obviously break down at high
optical depths, because the gas at high densities will typically be
shocked and follow a highly stochastic evolution, developing small-scale
structure. The spectrum will also be greatly affected by velocity
caustics and thermal broadening, and when $\tau \gg 1$ the value of
$\delta$ derived from Gaussianization will have large errors because of
saturation. On the other hand, at low densities the evolution is much
more regular, velocity caustics do not appear, and thermal broadening
can be neglected, so there should be a good correspondence between  the
optical depth at a given velocity and the gas density at the
corresponding point in space.

  The 2-point distribution function for a Gaussian field is
\begin{equation}
P_2(\delta_1,\delta_2, \dvv) = \frac{1} {2 \pi
\sqrt{\xi(0)^2-\xi(\dvv)^2}} \,
\exp\left[-\frac{1}{2}\frac{\xi(0)(\delta_1^2+\delta_2^2)-
2 \delta_1 \delta_2 \xi(\dvv)}{\xi(0)^2-\xi(\dvv)^2}\right] \, .
\label{tpf}
\end{equation}
This can be rewritten as
\begin{equation}
P_2(\delta_1,\delta_2, \dvv) = P(\delta_1)
\frac{1}{\sqrt{2 \pi \xi(0)\, (1-\txi(\dvv)^2})} \, 
\exp\left\{-\frac{1}{2}\frac{[\delta_2-\txi(\dvv)\delta_1]^2}
                {\xi(0)\, [1-\txi(\dvv)^2 ] } \right\} \, .
\label{tpf2}
\end{equation}
Thus, for every value of $\delta_1$, we can estimate $\txi(\dvv)$
from the distribution of $\delta_2$ conditional to the value of
$\delta_1$. For example, one could estimate $\txi(\dvv)$ from the
median of $\delta_2$ (or any other adequate percentile), and see how 
the
result depends on $\delta_1$. If the method works, $\txi$ should be
approximately independent of $\delta_1$ for low values of $\delta_1$.
Using instead $\xi(\dvv) = \, < \! \delta({\mathbf v})\,
\delta({\mathbf v} + \dvv ) \! > \, $ is equivalent to averaging the
estimate of $\xi$ over all values of $\delta_1$ with equal weights.

\subsection{The Effect of Peculiar Velocities on the Power Spectrum}

In their reconstruction of the linear power spectrum from \lya forest
lines, CWKH neglect the effect of peculiar velocities
on the shape of $P(k)$.  This effect arises in the 
conversion from the measured one-dimensional power spectrum $P_{1D}$
to the desired three-dimensional 
power spectrum $P_{3D}$.
While the shape of $P_{3D}$ is not affected by peculiar velocities
in linear theory (Kaiser 1987), $P_{1D}$ {\it is} affected as
shown by Kaiser \& Peacock (1991):
\begin{equation}
    P_{1D}(k_\pa)=\frac{1}{2\pi}\int_{k_\pa}^\infty dk k P_{3D}(k)
    (1+\beta k_\pa^2/k^2)^2,
    \label{1D3D}
\end{equation}
To demonstrate how this will affect the reconstruction of CWKH,
we substitute a specific
$P_{3D}(k)$ into equation (\ref{1D3D}), with various values of $\beta$.  
We use the cold dark matter power spectrum parameterization
of Bardeen \etal (1986), with the coefficients of Ma (1996) 
(see \S 3.1 below), and the parameters: $\Omega=1$, $h=0.5$, $n=1.0$.
We then assume that the resulting $P_{1D}$ is the one-dimensional
power spectrum for the \lya forest and attempt to reconstruct the
three-dimensional power spectrum ignoring peculiar velocities
(i.e., setting $\beta=0$).
This yields a new function $\tilde{P}_{3D}(k)$, different from the
correct $P_{3D}(k)$ because of the effect of peculiar velocities,
which is given by 
\begin{equation}
\tilde{P}_{3D}(k) = P_{3D}(k) (1+\beta)^2-
\int_k^\infty dk'\left[\frac{4 \beta}{k'} P_{3D}(k')
(1+\beta\frac{k^2}{k'^2})\right].
\end{equation}
The second term causes the change in the shape of $\tilde{P}_{3D}(k)$
relative to $P_{3D}(k)$.
\begin{figure}
\centerline {
\epsfxsize=4.7truein
\epsfbox[70 32 545 740]{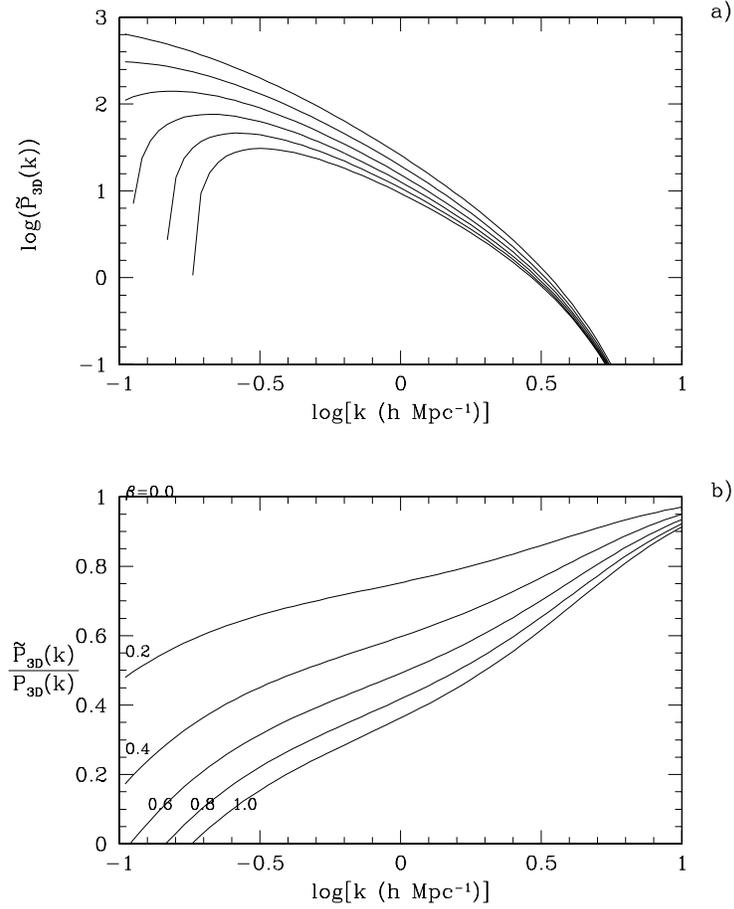}
}
\caption{(a) The power spectrum as it 
would appear if reconstructed from $P_{1D}(k)$ while ignoring 
peculiar velocity effects.  From top to bottom, 
$\beta=0.0, 0.2, 0.4, 0.6, 0.8, 1.0$.  The $\beta=0$ line 
is the correct power spectrum.
The normalization for all curves is fixed at large k,
and a Gaussian cutoff has been applied.
(b) The ratio of the reconstructed to the true power spectra. }
\label{f1D3D}
\end{figure}
Figure \ref{f1D3D} shows the results of this procedure for various
values of $\beta$.  The error in the reconstructed power spectrum
grows with scale and eventually causes the results to become 
negative. 
The power spectrum used in this figure was smoothed by a Gaussian 
with radius $r_s=0.24 h^{-1}\mpc$\footnote{
The value of $r_s$ quoted in CWKH (in the caption of their
Fig. 2) was wrong by a factor $2\pi$; 
Croft 1998, priv. communication.}.
CWKH found that this was necessary to match the $P(k)$ they
reconstructed from simulations.

  Thus, the shape of the mass power spectrum $P_{3D}(k)$ cannot be
recovered until the parameter $\beta$ is known. How can this parameter
be determined? One could in principle attempt to determine $\beta$
from observations in multiple lines of sight, as described in the
next section. The anisotropy of the correlation function depends on
the two parameters $\beta$ and $f(z)$. But as we shall see, a very
large amount of data will be required to measure both of them
independently. If $\beta$ can be predicted from theory, it should be
much less difficult to determine the power spectrum and $f(z)$.

  Linear theory would predict $b=2$ for a constant temperature in the
intergalactic medium, because the optical depth (which is the quantity
that is modified by peculiar velocities in the mapping from real to
redshift space) is proportional to the neutral hydrogen density, or
the square of the gas density. In a photoionized medium, the gas
temperature is determined by a balance of photoionization heating and
adiabatic cooling. The heating rate is proportional to the recombination
rate, $\alpha\rho$, and the adiabatic cooling is proportional to the
temperature $T$. Since $\alpha \propto T^{-0.7}$, the relation
$T\propto\rho^{0.6}$ is set up if the gas temperature is not affected
by shocks or by reionization (see Hui \& Gnedin 1997, Croft \etal
1997). This leads to a neutral hydrogen density proportional to
$\alpha\rho^2 \propto \rho^{1.6}$, and therefore a \lya forest bias
$b=1.6$. Since $\Omega(z) \simeq 1$ at $z\sim 3$ for viable models,
$\beta\simeq 0.6$. However, in reality the bias depends on the relation
of optical depth to the initial density on small scales, where
non-linearities are important, and therefore the correct bias to use
in equations (6), (9), and (10), where linear theory is applied to
large scales using the Gaussianization approximation, could be
substantially different. Numerical simulations can be used to calculate
a better value for the \lya forest bias (which will in general depend on
redshift), but of course this will only
be accurate if the simulations are modeling the structure of the \lya
forest correctly.

  The effects of peculiar velocities on the recovery of the mass power
spectrum have independently been pointed out in a recent paper by Hui
(1998), which appeared as this paper was being completed.

\section{Errors in the Measurement of the \lya Forest Correlation}

In this section, an estimate is obtained of the statistical error in
measuring the parameters of the correlation function, due to the random
nature of the absorption lines that appear in the spectra. A simple
model of absorption lines will be used to generate random spectra that
reproduce the characteristics of the individual \lya forest absorption
lines without the large scale correlation.

\subsection{Parameterization of the Correlation Function}

As discussed in \S 2, two parameters describe the anisotropy of
the correlation function:  $f(z)$, reflecting the effect of the 
cosmological geometry, and $\beta(z)$, incorporating the peculiar
velocity effects.  We also parameterize the shape of the power 
spectrum to a fitting formula for cold dark matter models,
\begin{equation}
P(k_s)=\frac{k_s^n [\ln(1+\alpha_1 q)/\alpha_1 q]^2}
         {[1+\alpha_2 q +(\alpha_3 q)^2+(\alpha_4 q)^3 +
	  (\alpha_5 q)^4]^{\frac{1}{2}}},
	  \label{pspec}
\end{equation}
where $q\equiv k_s/\Gamma(z)$. We have reexpressed the power spectrum in
terms of $k_s=k (1+z)/H(z)$. The free parameters of this model for the
shape of the power spectrum are $n$ and $\Gamma(z)$ 
(given by $\Gamma(z)=(1+z) \Omega_0 h^2 /H(z)$).  
The formula is given by Bardeen \etal (1986), but we modify the 
parameters to the fit for $\Omega_b=0.05$: 
$\alpha_1=2.205$, $\alpha_2=4.05$, $\alpha_3=18.3$, $\alpha_4=8.725$,
and $\alpha_5=8.0$ (Ma 1996).
In addition we include a smoothing parameter by multiplying the
above formula by the factor $\exp(-k_s^2 v_s^2/2)$, with
$v_s = r_s H(z)/ (1+z) = 48\kms$ at $z=3$.
This is motivated
by the result of CWKH, who find that the \lya forest has
a power spectrum (after Gaussianization) that can be well 
approximated by a smoothed version of the power spectrum of the mass.
We fix $v_s$ (rather than $r_s$) to the same value for all models, 
since the observations always yield separations in terms of velocity.
 
\begin{figure}
\centerline {
\epsfxsize=4.7truein
\epsfbox[70 32 545 740]{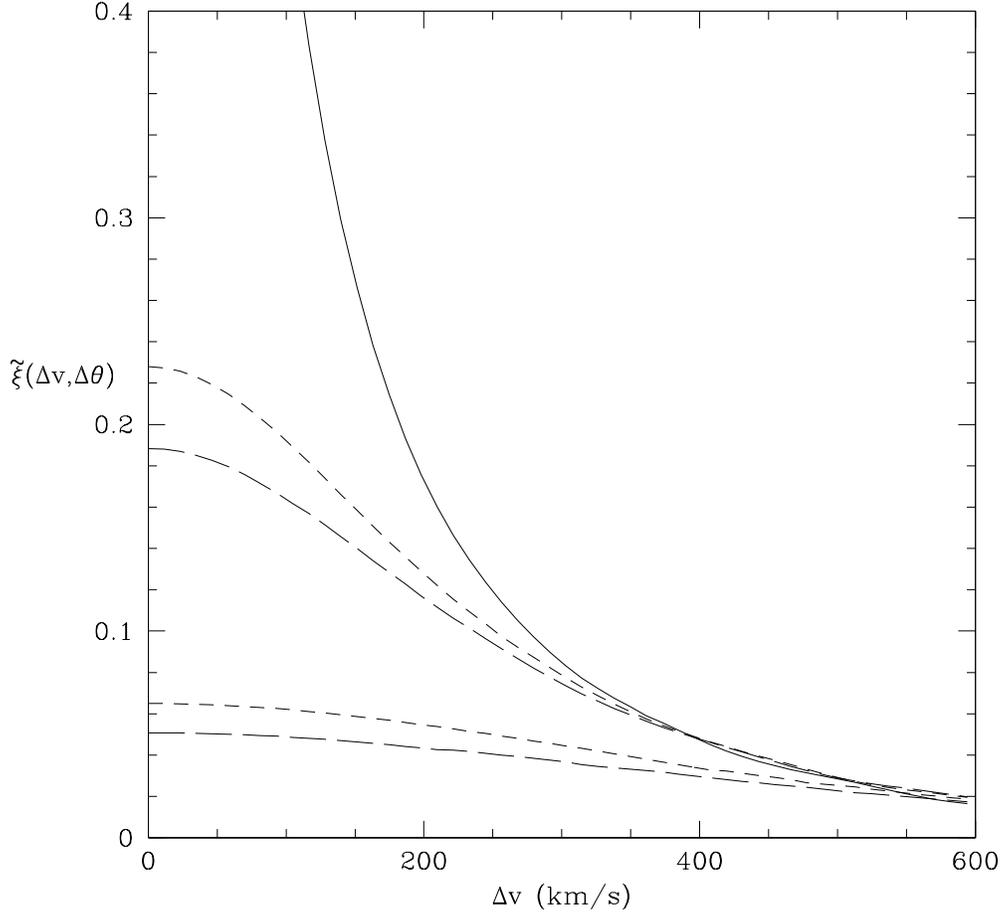}
}
\caption{The solid and long dashed lines are predictions for a flat
model with $\Omega_0=1.0$, $h=0.65$, and $\beta=0.6$ (giving 
$f(z)=1.61$). The solid line is the correlation along the line of
sight (normalized to $\tilde{\xi}(0,0)=1$
because of Gaussianization).
The upper dashed lines are for $\dt=127''$, and the
lower dashed lines are for $\dt=300''$.  The short dashed lines were
produced by changing $f(z)$ to 1.39, appropriate for a flat 
$\Omega_\Lambda=0.6$ model, {\it without} changing the power spectrum
parameters or $\beta$ (the line of sight correlation is the same in
each case).}
\label{pred1}
\end{figure}

In Figure \ref{pred1}, we display the computed $\txi(\dv, \dt)$
along the line of sight ($\dt = 0$), and at separations $\dt=127''$
and $\dt=300''$, at mean redshift $<z>=2.25$. Because of the
Gaussianization, we measure $\txi(\dv, \dt) \equiv
\xi(\dv, \dt)/\xi(0)$. The model parameters are $\Omega_0=1.0$,
$\Omega_\Lambda=0.0$, $h=0.65$, $n=1.0$, and $\beta=0.6$.
These parameters correspond to $f(z)=1.61$, and
$\Gamma(z)=0.0036\,  ({\rm km/s})^{-1}$.
Also shown in the figure are the predictions for a model
with $f(z)=1.39$ (the value for $\Omega_0=0.4$, $\Omega_{\Lambda}=0.6$),
but with $\Gamma$, $n$, $\beta$ and $v_s$ unchanged.
 
\subsection{Analysis of Random Spectra}

In this subsection we use a random line model to estimate the noise in
the measurement of $\txi(\dv, \dt)$. 
We create \lya forest
spectra using a code which produces lines by randomly distributing  
Voigt profiles with a specified distribution of column densities $N$ 
and widths $b$.   
\begin{figure}
\centerline {
\epsfxsize=4.7truein
\epsfbox[70 32 545 740]{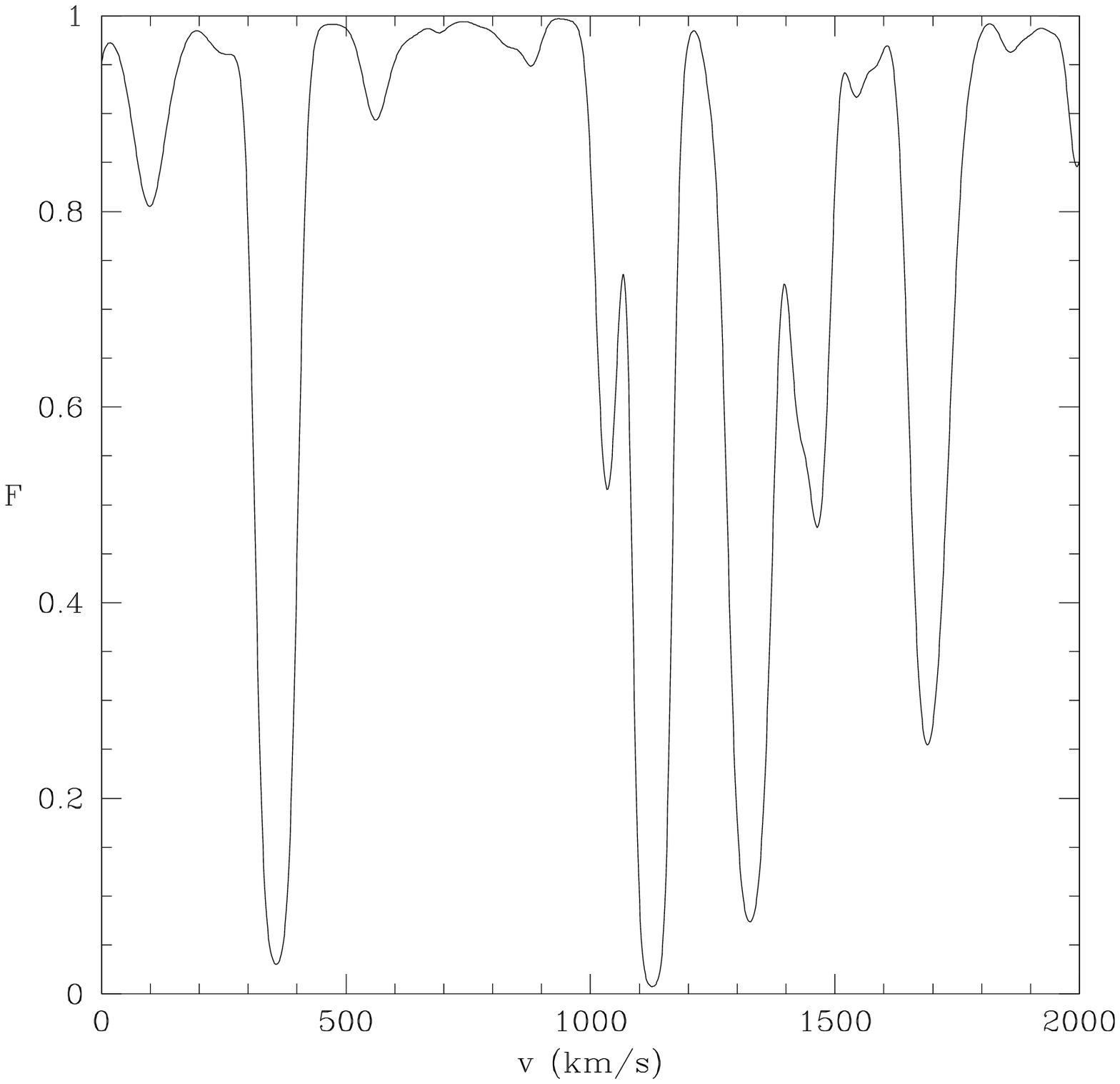}
}
\caption{ Flux vs. velocity for a 
randomly generated spectrum.}
\label{ranlines}
\end{figure}
Figure \ref{ranlines} shows a piece of one of these spectra.
We take a set of parameters 
that closely match the distribution in the observations
(e.g., Kim \etal 1997).  We will consider the 
mean redshift $<z>=2.25$.  We set the number 
of lines per unit redshift with column density greater than 
$10^{14}\cm^{-2}$ to be $N_{>14}=50$, and set $f(N)\propto N^{-1.35}$
with a break to $N^{-1.7}$ at $\log(N)=14.3$.  
We set the mean b-parameter 
of the Voigt
profiles to be $<b>=30\kms$ with a Gaussian dispersion of 
$\sigma_b=12\kms$ and
a lower cutoff of $b_{cut}=24\kms$.  

\begin{figure}
\centerline {
\epsfxsize=4.7truein
\epsfbox[70 32 545 740]{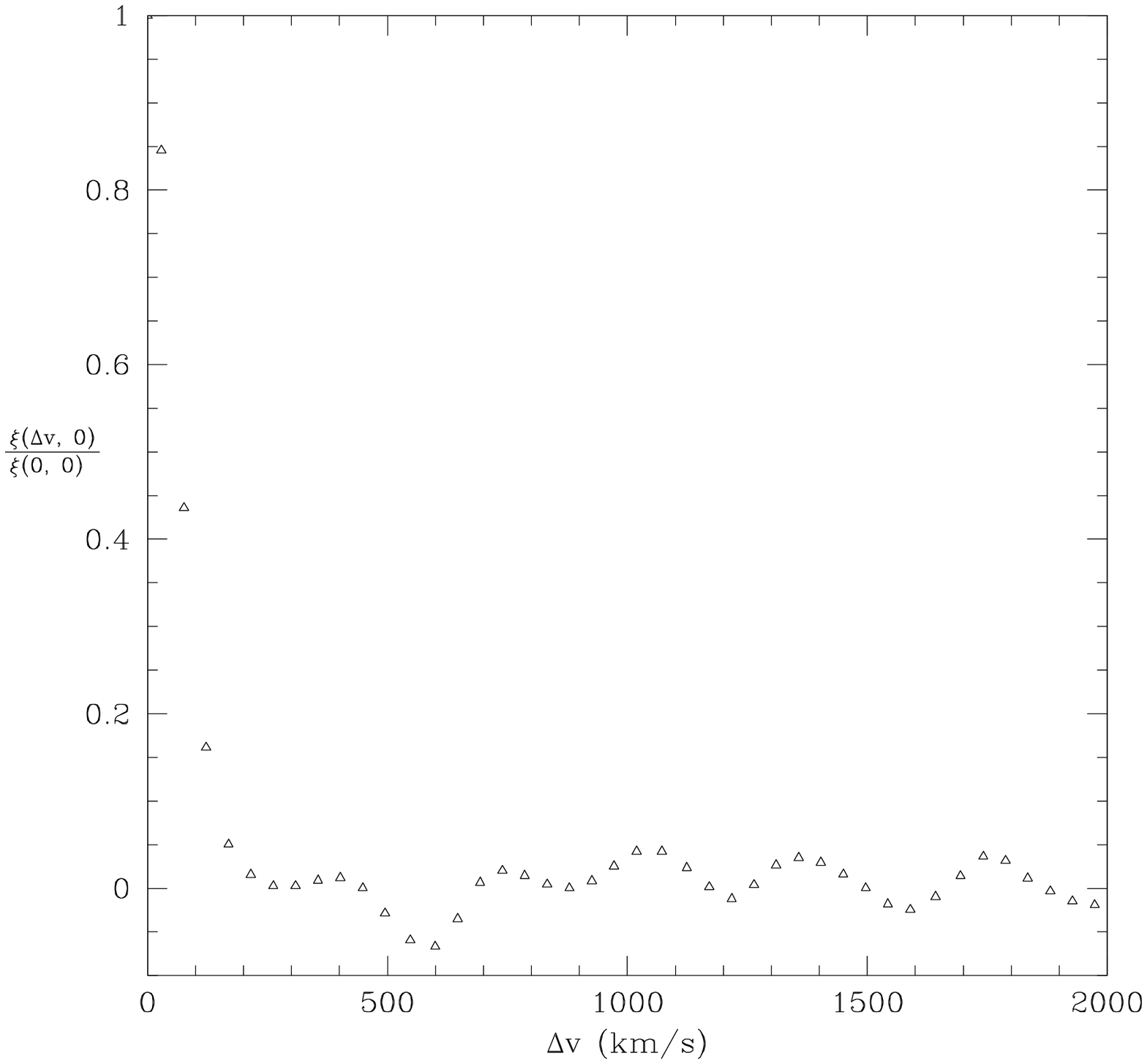}
}
\caption{ The correlation function along
the line of sight for a pair of randomly generated spectra
($\dz=0.5$, $<z>=2.25$ for each).}
\label{cor2fig}
\end{figure}

A spectrum is generated by calculating the transmitted flux at 
discrete pixels from the list of randomly generated lines.
We then make a transformation of the transmitted flux to a new
variable $\delta$, requiring that the probability distribution of
$\delta$ is Gaussian (this is the Gaussianization procedure). 
Because the absorption lines are random, 
there should be no correlation at separations beyond the width of the
individual components, so any correlation we measure at larger
separations is due to noise. Figure \ref{cor2fig} shows the
resulting $\txi(\dv, 0)$ for a pair of lines. 
The self-correlation of the individual components due to their own
width extends out to $\dv \simeq 170\kms$. 
The {\it rms} fluctuation around the
mean $\txi$ is $\sigma_\txi \simeq 0.03$.
The errors in $\xi$ are obviously also correlated over $\sim 100\kms$
due to the width of the lines.

\subsection{Estimation of Errors}

\begin{figure}
\centerline {
\epsfxsize=4.7truein
\epsfbox[70 32 545 740]{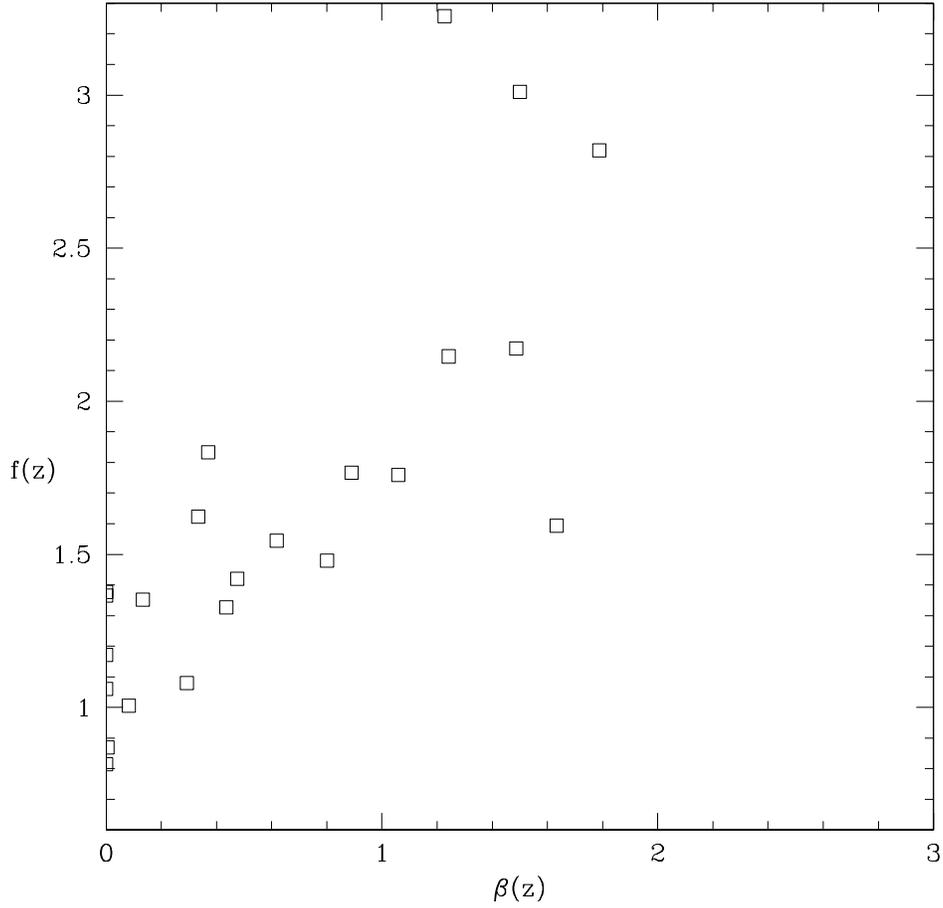}
}
\caption{Best fit values for $\beta(z)$ and 
$f(z)$ for 25 sets
of random spectra.
The sets of quasar spectra fit were random realizations of a triplet
with angular separations $127''$, $147''$, and $177''$, with a useful
redshift range $z=2.0-2.5$.
The true values were $\beta(z)=0.6$ and 
$f(z)=1.61$. 
$f(z)=1.61$ corresponds to $\Omega_0=1.0$ and
$\Omega_\Lambda=0.0$ at redshift $<z>=2.25$.
Two points with high $\beta$ do not
appear on the plot.  The constraint $\beta \geq 0$ was applied to
the fits.}
\label{fitplot}
\end{figure}

  We now estimate the errors in measuring parameters in the correlation
function from a given set of quasar spectra. 
We consider first as an example the
triplet of quasars of Crotts \& Fang (1998), which have a
useful redshift range $z \simeq 2.0-2.5$ and separations $\dt=127''$,
$147''$, and $177''$.  
The error bars we have
computed for $\txi$ are not necessarily Gaussian, and certainly not
independent, so we use a Monte Carlo technique. We introduce noise to
the calculated correlation function of a model by adding the noise
values of the correlation measured from spectra of randomly generated
lines (i.e., we sum the curves in Fig. 3 and 5). We then fit the five
parameters of the model, minimizing the $\chi^2$ of the correlation
function in linear bins of $\sim 45\kms$, 
using the error bars found from
the dispersion in the correlation obtained from the spectra of random
lines. We repeat this procedure independently 25 times and fit each 
of the 25
simulated correlation functions separately. This should give the range
of the fitted values we would expect to obtain from data with some
fixed true parameter values. We do not use the correlation function
at separations less than $300 \kms$, approximately the scale of
non-linearities.
Changing this restriction to $200\kms$ results in a small improvement
(particularly, tighter correlation between $f$ and $\beta$), but
does not substantially change our conclusions. 
Figure \ref{fitplot} shows the
results for the $f(z)=1.61$, $\beta=0.6$ model, fit to the triplet
of quasars described above.

\begin{figure}
\centerline {
\epsfxsize=4.7truein
\epsfbox[70 32 545 740]{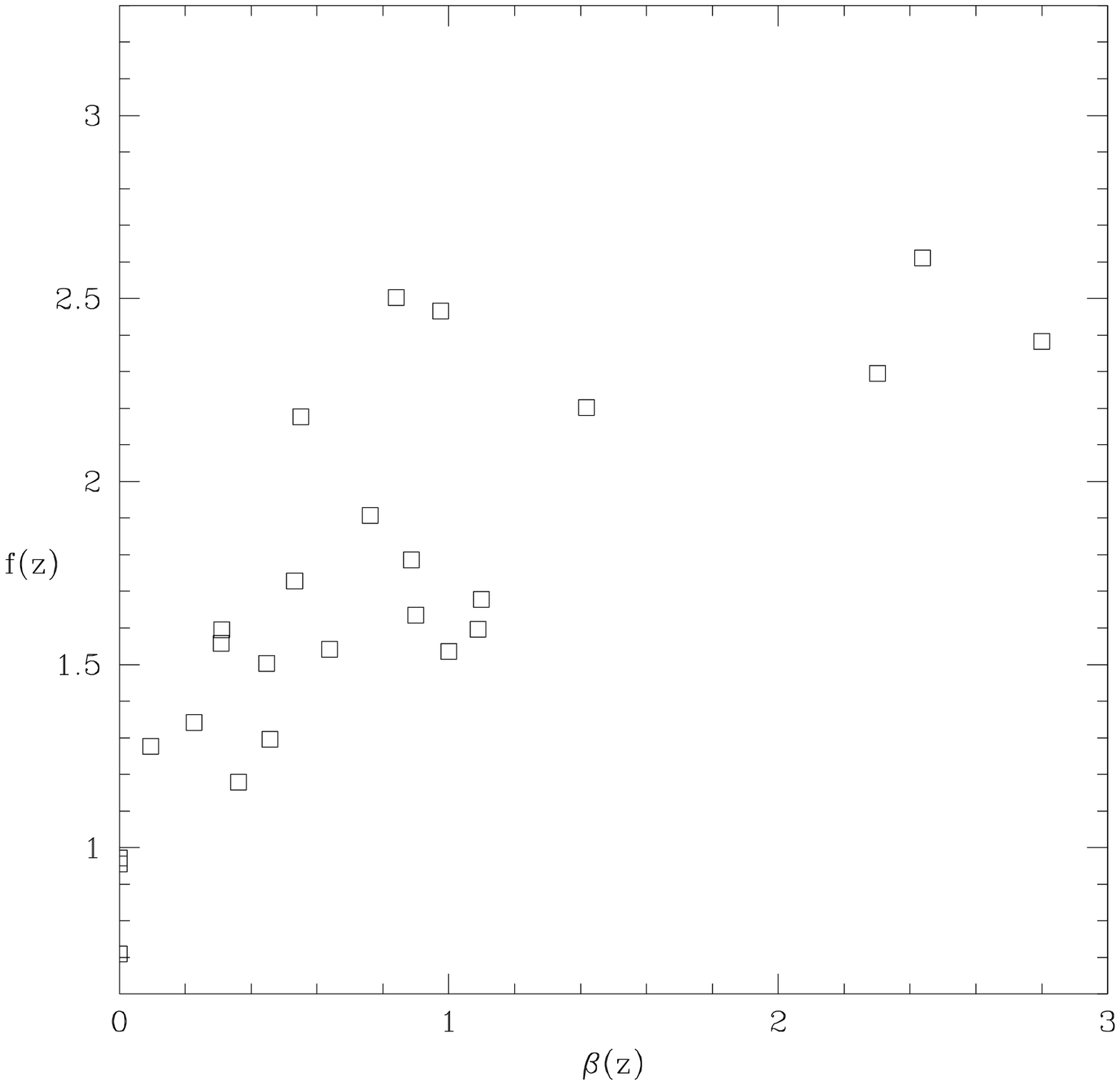}
}
\caption{Best fit values for 25 realizations of a combination of
6 pairs of quasars
separated by $\dt=45''$, $75''$, $105''$, $135''$, $165''$, and
$195''$.    
The true parameter values were $\beta(z)=0.6$ and $f(z)=1.6$
($\Omega_0=1.0$, $<z>=2.25$). 
The constraint $\beta \geq 0$ was applied to
the fits.}
\label{fitplot2}
\end{figure}

\begin{figure}
\centerline {
\epsfxsize=4.7truein
\epsfbox[70 32 545 740]{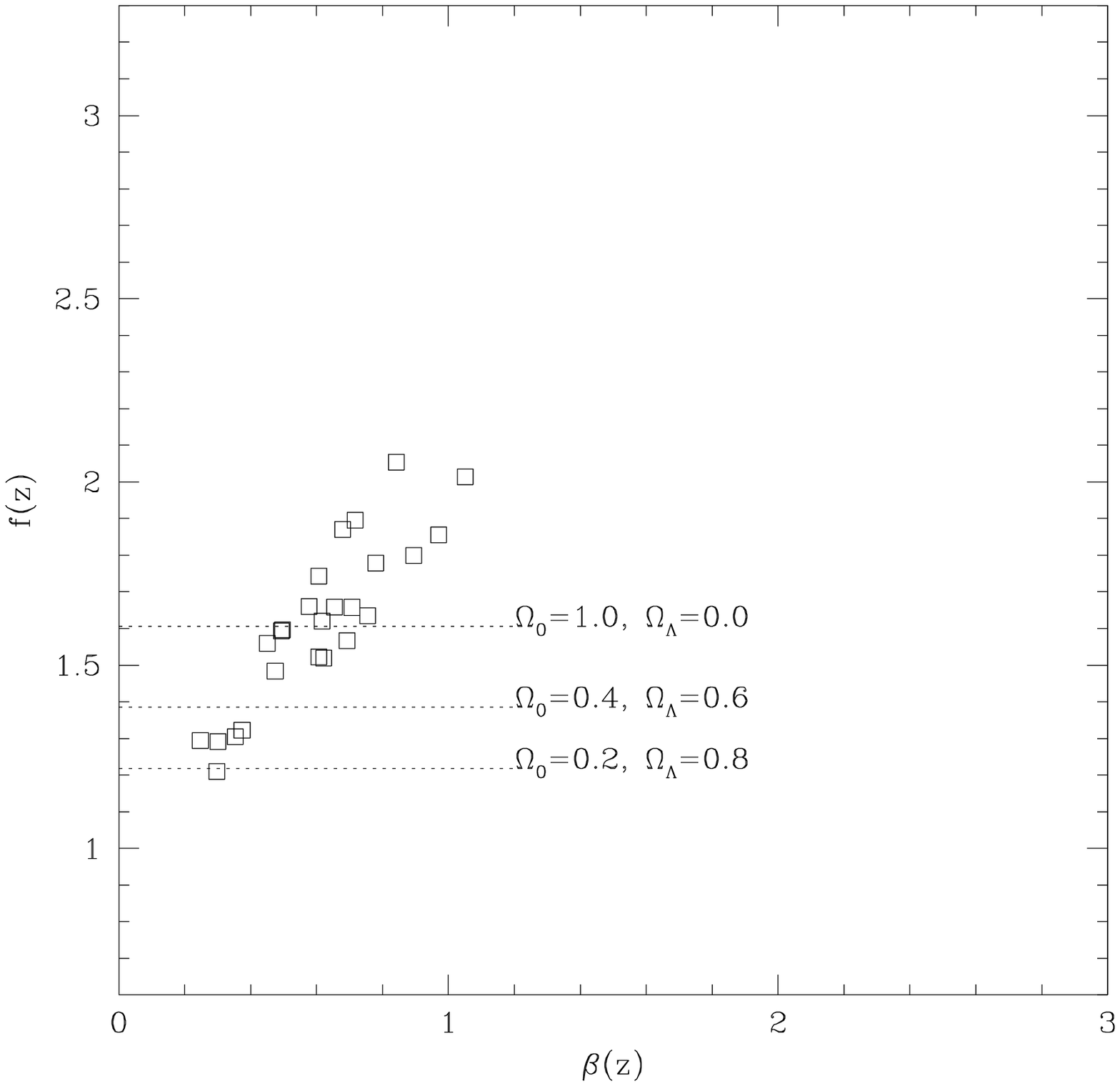}
}
\caption{This figure shows the same 25 fits as Fig. \ref{fitplot2},
except the added noise in the cross-correlation has been reduced by 
a factor of 2, and the added noise in the line of sight correlation
has been reduced by a factor of 6.}
\label{fitplot3}
\end{figure}

\begin{figure}
\centerline {
\epsfxsize=4.7truein
\epsfbox[70 32 545 740]{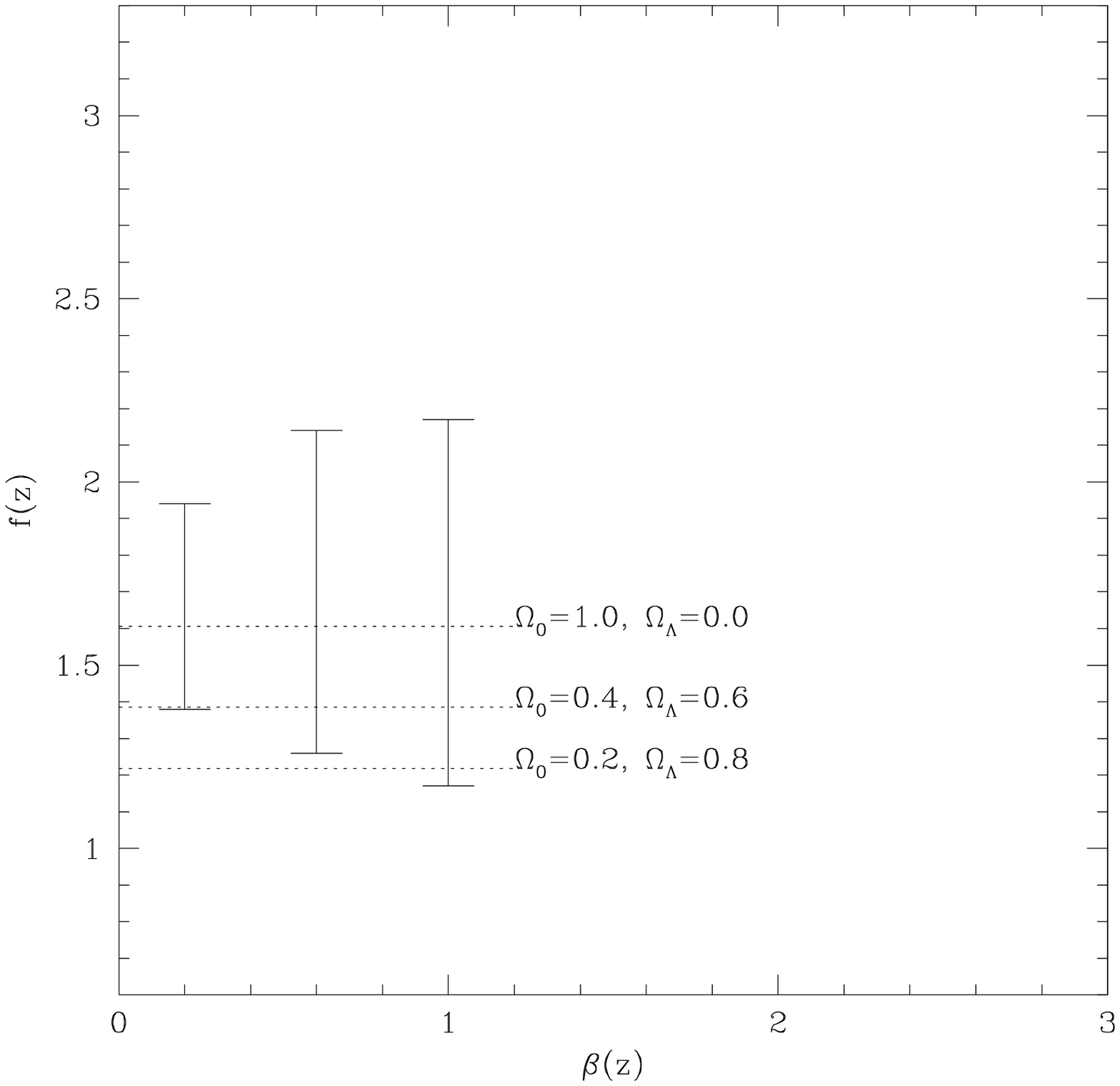}
}
\caption{Fits assuming $\beta$ is known ($\beta$ is held fixed at the
correct value).  The error bars represent 
approximately 92\% confidence.  
The 25 sets of quasars fit each contained
6 pairs as in Fig. \ref{fitplot2}.  The true value of $f$ is
$f=1.6$.  The error bars exclude one point at each extreme.  We
show $\beta=0.2$, $\beta=0.6$, and $\beta=1.0$.}
\label{fitplot4}
\end{figure}

To reduce the scatter, it is necessary to add more quasars.    
Figure \ref{fitplot2} shows
the improvement that can be expected by combining multiple pairs of 
quasars at different angular separations 
(but still at the same redshift).  Six pairs with separations
$\dt=45''$,  $75''$, $105''$, $135''$, $165''$ and $195''$, 
with useful redshift range $z=2.0-2.5$, are combined.
In Figure \ref{fitplot3}, we use the same 6 angular 
separations but reduce the noise in the cross-correlation
by a factor of two, and reduce the noise in the LOS correlation
by a factor of six.
This should approximately represent 24 pairs of quasars, with an
additional 384 single quasars included to improve the measurement
of the line of sight correlation. 
An independent determination of $\beta$ can also improve the 
measurement of $f$.  In Figure \ref{fitplot4} we plot the 
distribution of measured $f$ for $\beta=0.2$, $\beta=0.6$, and
$\beta=1.0$.  For these fits, the value of $\beta$ is fixed at
the correct value.  The sets of quasars used are the 6 pairs of
Figure \ref{fitplot2}.
Finally, with fixed $\beta$, we found that $f$ 
can be measured to $\sim \pm 10\%$ ($1\sigma$) using the reduced noise 
(24 quasar pairs) of Figure \ref{fitplot3}.

\section{Discussion}

  We have presented a method for analyzing the \lya
forest spectra along parallel lines of sight in order to extract
the geometric parameter $f(z)$ and the quantity determining the strength
of peculiar velocity distortions $\beta(z)\simeq\Omega^{0.6}/b$.
We have worked here in the context of the linear regime, where the
correlation function is small and the peculiar velocities generally
cause large-scale structures (with positive or negative density
fluctuations) to be flattened along the line of sight.
Using the Gaussianization technique of Croft \etal (1998), one can
measure the full redshift space correlation function and obtain
$\beta$ and $f(z)$. 
Using a model in which we distribute a realistic set of
discrete lines randomly in space we have estimated the statistical
errors expected for a measurement of the correlation function.
We estimate that dozens of pairs of quasars are needed to 
reliably distinguish competing cosmological models.

  The parameter $f$ should probably be easier to measure going to
smaller separations, in the non-linear regime. The amplitude of
the correlation is then much larger, so many fewer pairs of quasars
should be needed to measure the anisotropy of the correlation function
to a fixed relative accuracy, required to constrain $f$ to interesting
levels. For small separations, the effect of peculiar velocities
should be opposite to that in the linear regime: the correlation
function should be elongated along the line of sight. This is caused
by the well-known ``fingers of God'' effect in galaxy redshift
surveys (where high density clusters appear as highly elongated
filaments pointing toward us). In the \lya forest, the ``fingers of
God'' effect is simply the contribution of the internal velocity
dispersion (either hydrodynamic or thermal) of the absorbers to their
width.

  The disadvantage of working in the non-linear regime is that the
anisotropy of the correlation function can only be predicted with
numerical simulations; in addition, the 2-point joint probability
function is no longer given by equation (9), because the density
field is no longer Gaussian. Therefore, we can only measure $f$ if
we can be certain that numerical simulations are accurate enough.
But a more important reason to measure the 2-point function in the
non-linear regime is probably that it will provide a very powerful
test of the results of numerical simulations, and the large-scale
structure theories in which they are based. If the \lya forest
arises from individual, high-density clouds, the fingers of God
should be much more elongated and common than in the modern theories
of gravitational instability from primordial fluctuations.
Essentially, the full redshift space 2-point function gives a more
quantitative version of the measurement of transverse sizes of the
absorbers from coincidences of lines (Crotts \& Fang 1998 and references
therein).

  On the large scales, an alternative to finding a large number of
pairs of bright quasars to measure the anisotropy of the linear
correlation function to high accuracy could be to work with spectra
of much fainter, but much more numerous sources. So far, observational
studies of the \lya forest have used bright quasars as sources because
of the desired high resolution and signal-to-noise that is necessary
to measure the properties of individual absorption lines. But the
correlation function could in principle be obtained from spectra of
much poorer quality, as long as a very large number of spectra are
taken. Recently, large numbers of galaxies are being
identified at high redshift with the method of the Lyman limit break
technique (Steidel \etal 1996). Spectra are now being taken routinely
of galaxies in the redshift range $2.5$ to $4$, which have a number
density of $\sim $ one galaxy per square arc minute (Steidel \etal
1998). If the stellar
continuum of the galaxies can be modeled, the \lya forest spectra in
these galaxies could provide a better way to measure the \lya forest
correlation on large scales (in this case, Gaussianization could not
be used because the transmitted flux is measured only with very poor
resolution, so numerical simulations would be necessary to predict
the correlation function of the smoothed transmitted flux). These
observations could also be used to study the cross-correlation of
the \lya forest with the Lyman limit break galaxies.

  Future work is needed along two paths: analysis of numerical
simulations and observational data. CWKH showed that the
Gaussianization method works fairly well when applied to numerical
simulations. Direct application of the full method to recover
the redshift space correlation function to numerical simulations
should indicate the effect of non-linearities and give a more
accurate estimate of the accuracy that can be achieved in the
observational determinations. The effect of the finite size of the
numerical simulations should now also be more easily understood,
since the analytic method used in this paper predicts how the
correlation should behave on the large scales.

\end{document}